# Stochastic End-to-End Latency Modeling of the IoT-Edge-Cloud Continuum: Impact of Jitter and Traffic Variability on Deterministic Service Provisioning

K. Aghababaiyan, *Member, IEEE*, J. Gozalvez, *Fellow, IEEE*, and B. Coll-Perales, *Senior Member, IEEE*

***Abstract*—6G will integrate communication and computing capabilities in a IoT-edge–cloud continuum, enabling nodes to distribute workloads across the continuum. To support time-sensitive services, both communications and computing latencies must be controlled. Two key sources of temporal variability are arrival-time jitter and traffic variability. They can both impact the timing at which data is generated, transmitted and processed, and the resulting fluctuations can propagate throughout the continuum, increasing latency uncertainty. This paper studies the impact of stochastic temporal variability on the ability to support end-to-end deterministic service levels across the continuum. To this end, we present a novel queueing-based end-to-end latency model for the continuum, which we openly release. The model jointly captures computing and communication latency, and characterizes the complete end-to-end latency distribution, including tail latency. Our analysis shows that services with stringent latency deadlines and larger computing demands are more sensitive to temporal variabilities, making local execution the preferred option. In contrast, services with more relaxed deadlines are more resilient to temporal variabilities when executed locally or at the edge despite higher average and tail latencies. Edge offloading is beneficial under good cellular connectivity and increasing local processing workloads, whereas cloud execution is more sensitive to traffic variabilities because of the additional communication latency. Our analysis also shows that services offloaded are more sensitive to traffic variability than jitter due to higher communication latencies. These findings highlight that effective service offloading must jointly consider service requirements and sources of temporal variability to guarantee deterministic service levels.**

## I. Introduction

6G systems will seamlessly integrate communication and computing to support services across the IoT-edge-cloud (or local–edge–cloud) continuum. Through the continuum, devices can access additional and more powerful computing resources to meet growing computing demands. Much of this demand is driven by time-sensitive machine-type or cyber-physical services that require deterministic service levels, i.e. guarantees that a task will be completed before a bounded latency deadline. While significant research has focused on deterministic communications, future continuums must guarantee end-to-end deterministic performance, including both the communications and computing processes.

Deterministic performance cannot be guaranteed based on average latency alone. Instead, it requires bounding tail latency and controlling latency variation (jitter), since service performance is often influenced by rare but extreme latency events as well as dynamic variations in communications and computing loads. The ability to guarantee end-to-end deterministic service levels across the continuum is strongly influenced by stochastic temporal variabilities, particularly arrival-time jitter and traffic variability. Arrival-time jitter (or jitter in short) refers to random deviations of packet arrival times from their nominal schedule while preserving the average arrival rate, whereas traffic variability refers to temporal fluctuations of the packet arrival rate around its mean value, resulting in time-varying traffic load. Unlike arrival-time jitter, which perturbs packet arrival instants without changing the average arrival rate, traffic variability modifies the offered traffic load over time. These temporal variabilities affect the timing at which data is generated, transmitted and processed, and their effects can propagate throughout the continuum, increasing latency uncertainty, potentially leading latency deadline violations, and ultimately reducing the ability to guarantee end-to-end deterministic service levels. Conventional latency metrics and analytical models based on average values or static configurations fail to capture the stochastic effects. Instead, stochastic end-to-end latency models that jointly account for communication and computing processes are required to understand and accurately characterize the impact of stochastic temporal variabilities on deterministic service provisioning and to design effective service offloading policies for the continuum.

This paper advances the state of the art by introducing a novel queueing-based stochastic end-to-end latency model that jointly captures computing and communication latency across the entire IoT–edge–cloud continuum including the radio access network (RAN), transport network, core network, Internet, and computing infrastructure. The proposed model characterizes the end-to-end latency distribution under stochastic jitter and traffic variability, capturing latency variability and tail behavior through queueing-based analysis. The complete model is

K. Aghababaiyan is supported in part by the European Union under the 2024 MSCA Postdoctoral Fellowship program (project no. 101207598). This research has also been partially funded by MCIN/AEI/10.13039/501100011033 (PID2023-150308OB-I00), UMH, and Generalitat Valenciana (CIAICO/2024/167). (Corresponding author: Keyvan Aghababaiyan).

Keyvan Aghababaiyan, Javier Gozalvez and Baldomero Coll-Perales are with the Networked Systems Lab, Universidad Miguel Hernández de Elche, 03202 Elche, Spain (e-mail: kaghababaiyan@umh.es; j.gozalvez@umh.es; bcoll@umh.es).

openly released with this paper [1] providing the research community with a valuable tool to analyze how stochastic temporal variability propagates throughout the continuum and affects deterministic service provisioning.

Using the proposed model, we conduct the first comprehensive study of the impact of stochastic temporal variabilities on the ability to guarantee end-to-end deterministic service levels across the IoT–edge–cloud continuum. Specifically, we analyze the effects of arrival-time jitter and traffic variability. Through queueing-based stochastic analysis, we characterize the complete end-to-end latency distribution, including tail latency, and quantify the probability of meeting service latency deadlines under different communication and computing conditions when services are executed locally or offloaded to the edge or the cloud. Our analysis provides key insights into service offloading decisions and demonstrates that effective continuum management policies must jointly consider service requirements, communication and computing conditions, and potential stochastic temporal variabilities to reliably guarantee deterministic service levels.

The remainder of this paper is organized as follows. Section II reviews the state of the art. Section III presents system architecture. Section IV describes the proposed latency model for the computing components, and Section V presents the proposed latency model for all the communication components of the IoT-edge-cloud continuum. Section VI validates the proposed end-to-end latency model. Section VII outlines the evaluation scenarios, including the models used to represent arrival-time jitter and traffic variability. Section VIII and IX analyze the impact of arrival-time jitter and traffic variability, respectively, on the end-to-end latency and the ability to guarantee deterministic service levels across the continuum. Finally, Section X concludes the paper and summarizes the main findings.

## II. State of The Art

The objective of supporting latency-sensitive services in 5G and beyond networks has stimulated significant research efforts on latency analyzing and modeling. A large part of this work efforts has focused on characterizing the latency at the RAN given the impact of limited radio resources on latency under increasing network load. For example, [2] analyzes downlink latency using queueing theory, demonstrating how spectrum sharing can help guarantee strict latency requirements for Machine-Type Communication (MTC) services. In [3], the authors develop an analytical model to estimate the latency of 5G latency at the radio network level, accounting for different numerologies, modulation and coding schemes, and transmission schemes. Only a limited number of studies extend latency modelling beyond the RAN to also account for the transport and core networks latency. [4] presents an end-to-end 5G latency model for networked-based V2X communications that includes the radio, transport, core, and Internet latency. The model is subsequently used in [5] to analyze the ability of different 5G network deployments to support latency-sensitive vehicular services.

Studies analyzing the computing component of latency generally focus on task processing and queueing interactions, but do not account for complete end-to-end communication path. In [6], the authors present a queuing-based task allocation model for edge offloading, and jointly analyzed RAN latency, queueing delay and computing latency, although it is limited to average latency. In contrast, [7] presents a stochastic analytical model to evaluate queueing and edge computing latency and shows that computing latency can represent a significant fraction of the total latency in edge deployments. Similarly, [8] also investigates task offloading in time-sensitive edge computing environments using M/G/1 queues. The study quantifies average latency and its second-moment, but considers only computing latency and does not model RAN latency.

More recently, increasing attention has been devoted to the impact of stochastic temporal variability. The study in [9] derives latency violation probabilities and second-order latency moments at the RAN in the presence of jitter. In [10], the authors analyze latency jitter in industrial wireless networks, and optimize power allocation to minimize jitter. The study in [2] applies stochastic geometry and an M/G/1 queuing model to characterize base station behavior and analyze outage probability, average latency, and second-order latency moments as indicators of latency dispersion. In [11], the authors characterize the tail distributions of RAN latency and jitter, and analyze the delay violation probability to characterize the trade-off between delay and reliability. The study in [12] focuses on network-level determinism and proposes a deterministic latency/jitter-aware service function chain algorithm to guarantee bounded latency and near-zero packet loss. Reference [13] reviews latency management challenges across multi-domain networks and provides simplified analytical approximation alongside simulation-based evaluations to estimate latency bounds for network planning. Finally, expanding the scope beyond individual network domains, [14] investigates end-to-end latency and jitter guarantees across converged networks by modeling deterministic flow transmissions based on resource abstraction.

Collectively, these studies highlight the importance of deterministic service provisioning in future networks and advance towards a deeper understanding of how stochastic temporal variabilities like the jitter affect deterministic service provisioning. However, future IoT–edge–cloud continuums require a unified characterization of communication and computing latency, together with the stochastic temporal variability affecting both domains. Developing such an end-to-end stochastic model and using it to analyze deterministic service provisioning across the continuum are the primary objectives of this work.

## III. System Architecture and Model

This section presents the reference system architecture and the proposed end-to-end latency modelling approach for the IoT-edge-cloud continuum, which jointly models latency across both communication and computing domains. In this study, we consider the 5G system architecture depicted in Fig. 1(a), which follows 3GPP system architecture and the ITU-T reference architecture. A UE (User Equipment) includes local processors, and can access either edge or cloud computing nodes through the 5G network. To this end, the UE connects to a 5G base station (gNB) through the Uu interface of the Radio

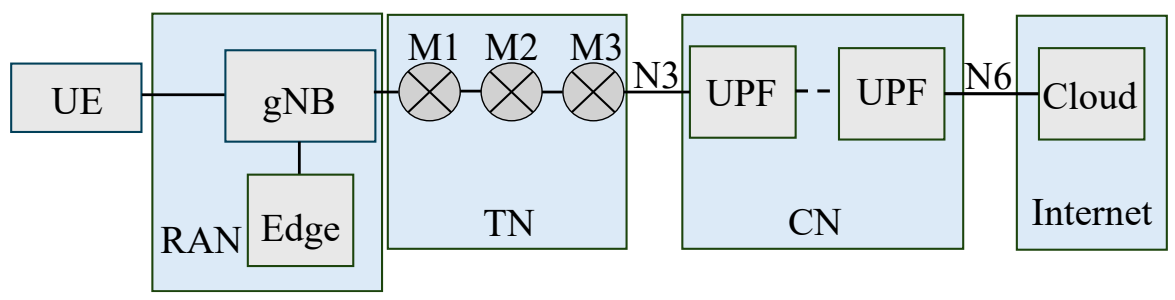

(a) 5G reference system architecture

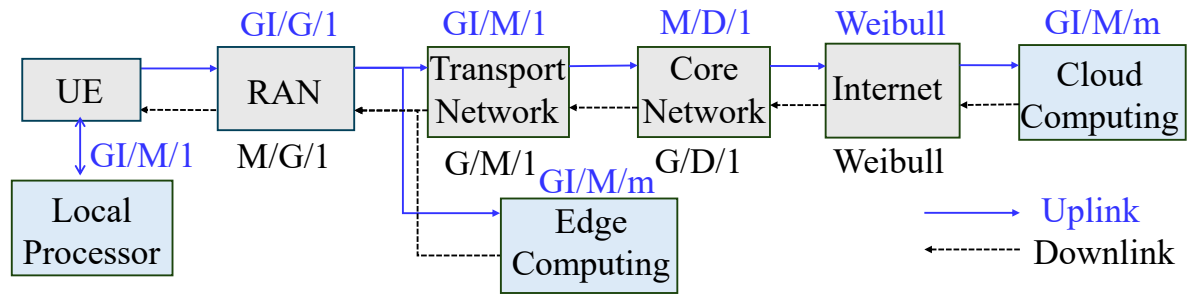

(b) End-to-end latency model

**Fig. 1.** System architecture and latency model.

Access Network (RAN). The generated data can then either be processed by an edge node connected to the gNB or traverse the Transport Network (TN) and Core Network (CN) to reach a cloud computing node. The TN interconnects the RAN and the CN through a set of multiplexing nodes deployed according to the hierarchical reference architecture defined by the ITU-T. Within the CN, traffic traverses one or more User Plane Functions (UPFs) before reaching the cloud. Edge nodes may also be connected to a transport network multiplexing node or directly to a UPF within the CN. In this work, however, we consider the edge deployment that minimizes end-to-end latency.

We model the end-to-end latency across the IoT–edge–cloud continuum using queuing theory following the reference architecture shown in Fig. 1(a). Fig. 1(b) depicts the queuing models adopted for each communication and computing component of the continuum. Different queueing models are considered for the uplink and downlink because the traffic characteristics and arrival processes differ in the two directions and depend on the previous component of the system architecture. In particular, the choice of each computing latency model depends on the arrival process and the number of cores assigned to each service. The choice of each communication latency model depends on the characteristics of the input traffic, including traffic load, the potential presence of jitter, and traffic variability of the communication component. All models represented in Fig. 1(b) are presented in detail in the following sections.

## IV. Computing Latency Models

### A. Local processing

The execution of tasks at the UE's local processor is modeled as a single-server queue, assuming that each application executes on a dedicated processor core. Packet arrivals are initially modeled as a Poisson process, while service times are assumed to follow an exponential probability distribution function (PDF). Under these assumptions, the local processing latency is naturally represented by an M/M/1 queue [15]. The service rate $\mu_l$ characterizes the available processor's computing capacity, where as the arrival rate $\lambda_l$ reflects the rate at which packets are generated. The system utilization factor is therefore $\rho_l = \lambda_l/\mu_l$ with the stability condition $\rho_l < 1$. The service time PDF is expressed as:

$$f_{T_{S_l}}(t) = \mu_l e^{-\mu_l t}, \quad (1)$$

where $\mathbb{E}[T_{S_l}] = 1/\mu_l$. In this case, the waiting time in queue, $T_{W_l}$, follows a mixed PDF:

$$f_{T_{W_l}}(t) = (1-\rho_l)\delta_0 + \rho_l.(\mu_l - \lambda_l).e^{-(\mu_l-\lambda_l)t}. \quad (2)$$

Consequently, a service is processed immediately ($T_{W_l} = 0$) with probability $1-\rho_l$, while, with probability $\rho_l$, it experiences an exponential waiting time with rate $\mu_l - \lambda_l$. The local processing latency is given by the sum $T_l = T_{W_l} + T_{S_l}$. Remarkably, in the M/M/1 case, this sum is again exponential with rate $\mu_l - \lambda_l$, a counterintuitive property of the Markovian system that greatly simplifies latency analysis [16]:

$$f_{T_l}(t) = (\mu_l - \lambda_l).e^{-(\mu_l-\lambda_l)t}. \quad (3)$$

The arrival process cannot be well approximated by exponential inter-arrival times when it exhibits variability and jitter. In such cases, the more general GI/M/1 queue model is more appropriate, as it accommodates general inter-arrival time distributions while retaining exponential service time. Under these conditions, inter-arrival times follow a general distribution with mean $1/\lambda_l$ and Laplace-Stieltjes Transform (LST) $\phi_{A_l}(s)$. The service time remains exponential with rate $\mu_l$, and the utilization is still $\rho_l = \lambda_l/\mu_l$. The waiting time PDF is no longer a simple exponential but instead is given by:

$$f_{T_{W_l}}(t) = (1-\beta)\delta(t) + \beta\mu_l(1-\beta)e^{-\mu_l(1-\beta)t}, t > 0, \quad (4)$$

where $\beta$ is the unique root of the following equation in the range (0,1):

$$\beta = \phi_{A_l}(\mu_l - \mu_l\beta). \quad (5)$$

The mean waiting time, based on classical GI/M/1 queueing theory, is obtained directly through $\beta$:

$$E[T_{W_l}] = \frac{\beta}{\mu_l(1-\beta)}. \quad (6)$$

The local processing latency based on the GI/M/1 queue model is then $T_l = T_{W_l} + T_{S_l}$, where the service time PDF is $f_{T_{S_l}}(t) = \mu_l e^{-\mu_l t}$ and $f_{T_l}(t)$ can be derived by the convolution:

$$f_{T_l}(t) = \int_0^t f_{T_{W_l}}(\tau) f_{T_{S_l}}(t-\tau)\, d\tau. \quad (7)$$

Evaluating the convolution for the Dirac delta component in (4) yields:

$$(1-\beta)\delta(t) * f_{T_{S_l}}(t) = (1-\beta)\mu_l e^{-\mu_l t}. \quad (8)$$

By calculating the exponential part of the convolution, we have

$$\beta\mu_l(1-\beta)e^{-\mu_l(1-\beta)t} * f_{T_{S_l}}(t) = \mu_l(1-\beta)e^{-\mu_l(1-\beta)t} - \mu_l(1-\beta)e^{-\mu_l t}, \quad (9)$$

and adding it to (8), the local processing latency simplifies to a unified exponential PDF with rate $\mu_l(1-\beta)$:

$$f_{T_l}(t) = \mu_l(1-\beta)e^{-\mu_l(1-\beta)t}. \quad (10)$$

### B. Edge Computing

An edge node typically comprises multiple processors (or cores) that can be dynamically allocated to execute tasks. This architecture is naturally modeled by a multi-server queue, where each processor operates as an independent parallel server. Under the common assumption of exponentially distributed service times, the system can be modeled as an M/M/m queue [17]. Since an edge platform receives traffic generated independently by multiple users (number of UEs is high) served by different base stations, the Palm–Khintchine theorem supports approximating the combined arrival stream as Poisson [16]. This is the case because when a large number of independent sources, each generating arrivals in a random but consistent manner, are combined, their superposition tends to exhibit Poisson-like behavior. This provides a sound justification for the M/M/$m$ model selection. In this model, $m = m_e$ denotes the number of cores, $\mu_e$ is the service rate of each allocated core, and $\lambda_e$ is the total arrival rate from different independent traffic sources. The resulting server utilization is $\rho_e = \lambda_e / m_e \mu_e$, and we assume the system operates under First-Come, First-Served (FCFS) scheduling. In steady state, the probability of zero services in the system is:

$$P_0^e = \left[ \sum_{k=0}^{m_e-1} \frac{\left(\frac{\lambda_e}{\mu_e}\right)^k}{k!} + \frac{\left(\frac{\lambda_e}{\mu_e}\right)^{m_e}}{m_e!} \cdot \frac{1}{1-\rho_e} \right]^{-1}, \tag{11}$$

and the expected number of services in the queue is given by the Erlang-C formula:

$$L_q^e = \frac{P_0^e \cdot \left(\frac{\lambda_e}{\mu_e}\right)^{m_e} \cdot \rho_e}{m_e! \, (1-\rho_e)^2}. \tag{12}$$

The average waiting time at an edge node is then $\mathrm{E}[T_{W_e}] = L_q^e / \lambda_e$. Under moderate and high loads, the waiting time distribution can be approximated by an exponential PDF [16]:

$$f_{T_{W_e}}(t) \approx \frac{1}{\mathrm{E}[T_{W_e}]} e^{-\frac{t}{\mathrm{E}[T_{W_e}]}}. \tag{13}$$

Under low load conditions, the waiting time in the edge node is negligible, i.e., $T_{W_e} \approx 0$. The service time for an edge processor is also exponential as:

$$f_{T_{S_e}}(t) = \mu_e e^{-\mu_e t}. \tag{14}$$

Hence, the total computing latency at the edge, $T_e = T_{W_e} + T_{S_e}$, becomes the convolution of two exponentials with different rates $f_{T_{W_e}}(t) * f_{T_{S_e}}(t)$, yielding a hypo-exponential PDF:

$$f_{T_e}(t) = \frac{\alpha_{W_e} \mu_e}{\mu_e - \alpha_{W_e}} \left(e^{-\alpha_{W_e} t} - e^{-\mu_e t}\right), \tag{15}$$

where $\alpha_{W_e} = 1/\mathrm{E}[T_{W_e}]$.

When the number of UEs offloading tasks to an edge node is small, the input process may deviate from Poisson. In such cases, the more general GI/M/m queue model is adopted. Under this model, arrivals are characterized by a mean $1/\lambda_e$, variance $\sigma_{A_e}^2$, and coefficient of variation $C_{A_e}^2 = \lambda_e^2 \sigma_{A_e}^2$. Using the well-known approximation for general arrival processes, the Erlang-C extension for GI/M/m modifies the mean queue length as:

$$L_q^e(\mathrm{GI/M/m}) \approx L_q^e(\mathrm{M/M/m}) \cdot \frac{1 + C_{A_e}^2}{2}, \tag{16}$$

where $L_q^e(\mathrm{M/M/m})$ is defined in (12). Accordingly, the mean waiting time for GI/M/m is $\mathrm{E}[T_{W_e}] = L_q^e(\mathrm{GI/M/m})/\lambda_e$, and it can be approximated as:

$$\mathrm{E}[T_{W_e}] \approx \frac{1 + C_{A_e}^2}{2} \cdot \frac{\left(\frac{\lambda_e}{\mu_e}\right)^{m_e} \rho_e}{\lambda_e m_e! \, (1-\rho_e)^2} P_0^e(\mathrm{M/M/m}). \tag{17}$$

The waiting time distribution can be approximated by an exponential PDF like (13). Then, the service time PDF remains exponential like (14), and the edge computing latency is obtained as the convolution of $f_{T_{W_e}}(t) * f_{T_{S_e}}(t)$ following (15) but using (17) for $\mathrm{E}[T_{W_e}]$.

*C. Cloud Computing:*

A cloud computing node relies on a multi-core architecture where individual cores operate independently. We model the cloud computing node as a GI/M/m queue, where $m = m_c$ represents the number of cores, and each core is characterized by a service time with an exponential PDF of rate $\mu_c$. The input traffic arrives from independent sources with average rate $\lambda_c$. The traffic arrives to the cloud node over an Internet connection. Consequently, the inter-arrival times exhibit a variability that can be accurately described by a Weibull distribution (See Section V.D) as it captures the variability of large-scale Internet traffic [18]. The GI/M/m model reflects a system with different processors operating under exponential service time assumptions, while accommodating the general distribution of arrivals. The number of cores $m_c$ at the cloud is larger than at the edge. The system utilization ratio is $\rho_c = \lambda_c / m_c \mu_c$. The mean waiting time $\mathrm{E}[T_{W_c}]$ is equal to $L_q^c(\mathrm{GI/M/m})/\lambda_c$. Based on the Erlang-C extension to GI/M/m, the mean queue length is given by:

$$L_q^c(\mathrm{GI/M/m}) \approx L_q^c(\mathrm{M/M/m}) \cdot \frac{1 + C_{A_c}^2}{2}, \tag{18}$$

$L_q^c(\mathrm{M/M/m})$ can be computed based on (12) replacing $\lambda_e$, $\mu_e$, $m_e$, $\rho_e$ with $\lambda_c$, $\mu_c$, $m_c$, $\rho_c$, respectively, and replacing $P_0^e$ with $P_0^c(\mathrm{M/M/m})$ which is the probability of zero services in the system in an M/M/m model. These changes result in:

$$L_q^c(\mathrm{GI/M/m}) \approx \frac{1 + C_{A_c}^2}{2} \cdot \frac{\left(\frac{\lambda_c}{\mu_c}\right)^{m_c} \rho_c}{m_c! \, (1-\rho_c)^2} P_0^c(\mathrm{M/M/m}), \tag{19}$$

where arrivals are characterized by $1/\lambda_c$, variance $\sigma_{A_c}^2$, and coefficient of variation $C_{A_c}^2 = \lambda_c \sigma_{A_c}^2$. The waiting time PDF $f_{T_{W_c}}$ can then be approximated by an exponential PDF under moderate and high load: $f_{T_{W_c}}(t) \approx \frac{1}{\mathrm{E}[T_{W_c}]} e^{-\frac{t}{\mathrm{E}[T_{W_c}]}}, t \geq 0$. Under low load conditions, the waiting time in the cloud node is negligible, i.e., $T_{W_c} \approx 0$. The service time for a cloud processor $f_{S_c}(t)$ remains exponential and is equal to $\mu_c e^{-\mu_c t}$. The distribution of the cloud computing latency ($T_c = T_{W_c} + T_{S_c}$) is derived as the convolution of $f_{T_{W_c}}(t) * f_{T_{S_c}}(t)$:

$$f_{T_c}(t) = \frac{\alpha_{W_c} \mu_c}{\mu_c - \alpha_{W_c}} \left(e^{-\alpha_{W_c} t} - e^{-\mu_c t}\right), \tag{20}$$

where $\alpha_{W_c} = 1/\mathrm{E}[T_{W_c}]$.

## V. Communication Latency Models

*A. Radio Access Network*

The RAN latency $T_{RAN}$ at the RAN consists of the transmission time (service time) and the waiting time: $T_{RAN} = T_S^{RAN} + T_W^{RAN}$. To compute the RAN latency, we assume a 5G RAN based on OFDM where the data rate per user depends on the instantaneous channel quality. Since small-scale fading generally dominates, the channel fading is commonly modeled as Rayleigh fading. Under Rayleigh fading, the instantaneous signal-to-noise ratio (SNR) $\gamma$ is exponentially distributed with mean $\bar{\gamma}$. According to the Shannon theory, the achievable communication rate $R_{RAN}$ can be expressed as

$$R_{RAN} \sim BW.\log_2(1+\gamma). \tag{21}$$

By normalizing with respect to the bandwidth $(BW)$, we obtain: $\widetilde{R_{RAN}} \sim \log_2(1+\gamma)$. For a packet of size $L$, the transmission time $T_S^{RAN}$ is equal to $\frac{L}{R_{RAN}}$, and the normalized transmission time is $\widetilde{T_S^{RAN}} = \frac{L}{\widetilde{R_{RAN}}} = BW.T_S^{RAN}$. The PDF of $\widetilde{R_{RAN}}$ is then:

$$f_{\widetilde{R_{RAN}}}(r) = \frac{\ln(2).2^r}{\bar{\gamma}} exp\left(-\frac{2^r-1}{\bar{\gamma}}\right), \qquad r \geq 0. \tag{22}$$

Since $\widetilde{T_S^{RAN}} = \frac{L}{\widetilde{R_{RAN}}}$, we use a transformation of variables to derive $f_{\widetilde{T_S^{RAN}}}(t)$. Using the Jacobian $\left|\frac{d\widetilde{R_{RAN}}}{d\widetilde{T_S^{RAN}}}\right| = \frac{L}{\left(\widetilde{T_S^{RAN}}\right)^2}$, we can derive the PDF of $\widetilde{T_S^{RAN}}$ as:

$$f_{\widetilde{T_S^{RAN}}}(t) = f_{\widetilde{R_{RAN}}}\left(\frac{L}{t}\right).\frac{L}{t^2}, \quad t > 0. \tag{23}$$

Substituting the expression for $f_{\widetilde{R_{RAN}}}(r)$, we obtain the closed-form PDF of the transmission time as:

$$f_{\widetilde{T_S^{RAN}}}(t) = \frac{\ln(2).2^{\frac{L}{t}}}{\bar{\gamma}} exp\left(-\frac{2^{\frac{L}{t}}-1}{\bar{\gamma}}\right).\frac{L}{t^2}, t > 0. \tag{24}$$

This expression captures the distribution of transmission time (service time) under Rayleigh fading. The result is heavy tailed, with higher variance under poor channel conditions (i.e. low $\bar{\gamma}$). To estimate the PDF $f_{T_W^{RAN}}(t)$ of the waiting time, we model the RAN using queueing theory and consider either an M/G/1 or a GI/G/1 queue depending on the variability at the arrival process. We first consider an M/G/1 queue, in which are arrivals approximated by a Poisson process and the service time follows a general distribution determined by wireless channel dynamics ([19],[20]). This model is well suited for the uplink when there is no variation in the input, as well as for the downlink where the input traffic arrives from an edge node with exponential service time or from the transport network when processing is done in the cloud. Since these preceding nodes have exponential service times, the traffic arriving at the RAN naturally follows a Poisson process, explaining the 'M' in the M/G/1 model. If the input uplink traffic varies or experiences jitter, the RAN is instead modeled as a GI/G/1 queue. In such cases, the inter-arrival times do not follow a Poisson process, making a general arrival distribution ('GI') necessary.

To derive $f_{T_W^{RAN}}(t)$ for the M/G/1 model, we move to the Laplace domain, where the Pollaczek–Khinchine formula provides an exact relation between the LST of the waiting time and the LST of the transmission or service time as:

$$\mathcal{L}_{T_W^{RAN}}(\theta) = \frac{(1-\rho)\theta}{\theta - \lambda + \lambda\mathcal{L}_{T_S^{RAN}}(\theta)}, \tag{25}$$

where $\rho$ is the utilization factor of RAN, $\lambda$ is the average arrival rate, and $\theta$ represents the Laplace transform variable. By normalizing the waiting time with respect to the bandwidth BW, we obtain $T_W^{RAN} = \frac{1}{BW}.\widetilde{T_W^{RAN}}$, $\mathcal{L}_{T_W^{RAN}}(\theta) = \mathcal{L}_{\widetilde{T_W^{RAN}}}(BW.\theta)$, and

$$\mathcal{L}_{\widetilde{T_W^{RAN}}}(\theta) = \frac{(1-\tilde{\rho})\theta}{\theta - \tilde{\lambda} + \tilde{\lambda}\mathcal{L}_{\widetilde{T_S^{RAN}}}(\theta)}, \tag{26}$$

where $\tilde{\rho} = \tilde{\lambda}.\mathrm{E}\left(\widetilde{T_S^{RAN}}\right)$ and $\tilde{\lambda} = \frac{\lambda}{BW}$. Based on $\widetilde{T_S^{RAN}} = \frac{L}{\widetilde{R_{RAN}}}$ and (21), the value of $\mathrm{E}\left(\widetilde{T_S^{RAN}}\right)$ is:

$$\mathrm{E}\left(\widetilde{T_S^{RAN}}\right) = L.\int_0^\infty \frac{1}{\log_2(1+\gamma)} f_\gamma(\gamma)d\gamma, \tag{27}$$

where $f_\gamma(\gamma) = \frac{1}{\bar{\gamma}} e^{-\frac{\gamma}{\bar{\gamma}}}$. We define $u$ as $u = \gamma + 1$, and $du = d\gamma$. Thus, we could simplify (27) as

$$\mathrm{E}\left(\widetilde{T_S^{RAN}}\right) = \frac{L\ln(2)\, e^{\frac{1}{\bar{\gamma}}}}{\bar{\gamma}} \int_1^\infty \frac{1}{\ln(u)} e^{-\frac{u}{\bar{\gamma}}} du. \tag{28}$$

To evaluate (26), the LST of $f_{\widetilde{T_S^{RAN}}}(t)$ is derived as

$$\mathcal{L}_{\widetilde{T_S^{RAN}}}(\theta) = \int_0^\infty e^{-\theta t} f_{\widetilde{T_S^{RAN}}}(t)dt. \tag{29}$$

By defining $x = \frac{L}{t}$, we have $t = \frac{L}{x}$ and $dt = -\frac{L}{x^2}dx$, and we can compute $\mathcal{L}_{\widetilde{T_S^{RAN}}}(\theta)$ in (29) as:

$$\mathcal{L}_{\widetilde{T_S^{RAN}}}(\theta) = \int_0^\infty e^{-\theta\frac{L}{x}} f_{\widetilde{T_S^{RAN}}}\left(\frac{L}{x}\right).\frac{L}{x^2}dx. \tag{30}$$

By replacing $f_{\widetilde{T_S^{RAN}}}(.)$, we could derive:

$$\mathcal{L}_{\widetilde{T_S^{RAN}}}(\theta) = \frac{\ln(2)}{\bar{\gamma}} e^{\frac{1}{\bar{\gamma}}} \int_0^\infty 2^x exp\left(-\frac{2^x}{\bar{\gamma}}\right) exp\left(-\frac{\theta L}{x}\right)dx. \tag{31}$$

Using the transformation $y = 2^x$, (31) can be computed as:

$$\mathcal{L}_{\widetilde{T_S^{RAN}}}(\theta) = \frac{e^{\frac{1}{\bar{\gamma}}}}{\bar{\gamma}} \int_1^\infty exp\left(-\frac{y}{\bar{\gamma}}\right) exp\left(-\frac{\theta L\ln(2)}{\ln(y)}\right)dy. \tag{32}$$

Once $\mathcal{L}_{\widetilde{T_W^{RAN}}}(\theta)$ is computed using (25), and considering $\mathcal{L}_{T_W^{RAN}}(\theta) = \mathcal{L}_{\widetilde{T_W^{RAN}}}(BW.\theta)$, the final step is to recover $f_{T_W^{RAN}}(t)$ by performing a numerical inverse Laplace transform. Finally, the PDF of the RAN latency is derived as $f_{T_{RAN}}(t) = f_{T_S^{RAN}}(t) * f_{T_W^{RAN}}(t)$.

For uplink with variable input traffic, the waiting time is better modelled with a GI/G/1 queue accommodating general independent inter-arrival times. Like the M/G/1 case, there is no exact closed-form expression for the LST of the waiting time distribution. However, under moderate and heavy loads, the mean waiting time $\mathrm{E}[T_W^{RAN}]$ can be accurately estimated using Kingman's approximation as:

$$\mathrm{E}[T_W^{RAN}] \approx \left(\frac{\tilde{\rho}}{1-\tilde{\rho}}\right)\left(\frac{C_A^2 + C_S^2}{2}\right)\mathrm{E}[T_S^{RAN}], \tag{33}$$

where $C_A^2$ and $C_S^2$ are the squared coefficients of variation for the inter-arrival and transmission times (service times), respectively. Consequently, the waiting time distribution in this generalized case can be approximated by an exponential PDF:

$$f_{T_W^{RAN}}(t) \approx \frac{1}{\mathrm{E}[T_W^{RAN}]} e^{-\frac{t}{\mathrm{E}[T_W^{RAN}]}}, t \geq 0. \tag{34}$$

Under low load conditions, the waiting time in the RAN is negligible, i.e., $T_W^{RAN} \approx 0$. The PDF of the RAN latency is derived through the convolution in the time domain $f_{T_{RAN}}(t) = f_{T_S^{RAN}}(t) * f_{T_W^{RAN}}(t)$.

### *B. Transport network*

The transport network (TN) latency $(l_{TN})$ is the sum of the propagation latency $l_{prop}^{TN}$ and the transit latency $T_t$. The propagation latency is the time it takes a packet to travel through the links interconnecting the TN nodes, and depends on the total distance $d_{TN}$ and the propagation speed $v_{TN}$. The transit latency is the time that a packet spends on the TN links, and includes both the transmission and queuing times.

As illustrated in Fig. 1, uplink traffic traverses $M_{n_{NT}}$ multiplexing nodes before reaching the core network. Accordingly, the transit latency is modeled as a $n_{NT}$ series of M/M/1 queues. Arrival times are approximated by a Poisson process with average rate $\lambda_{n_{NT}}^{TN}$, while service times are exponential with rate $\mu_{n_{NT}}^{TN}$. Although the traffic generated by individual gNBs (input to the TN) is not strictly Poisson, the Palm–Khintchine Theorem shows that aggregating many independent flows produces an overall arrival process that closely follows a Poisson distribution, thereby justifying the M/M/1 assumption. The resulting mean transport latency can then be expressed as:

$$\mathrm{E}[l_{TN}] = l_{prop}^{TN} + \mathrm{E}[T_t] = \frac{d_{TN}}{v_{TN}} + \sum_{i=1}^{n_{NT}} \left( \frac{1}{\mu_i^{TN} - \lambda_i^{TN}} \right). \tag{35}$$

The distribution of the transport latency is then:

$$l_{TN} = \frac{d_{TN}}{v_{TN}} + Exp(\mu_1^{TN} - \lambda_1^{TN}) * \dots * Exp(\mu_{n_{NT}}^{TN} - \lambda_{n_{NT}}^{TN}), \tag{36}$$

where $\mu_i^{TN}$ is the transmission rate (service rate) for each service over the link connecting multiplexing node $i$ with previous node, which can be expressed as:

$$\mu_i^{TN} = \alpha_i^{TN} . C_i^{TN}, \tag{37}$$

where $C_i^{TN}$ is the data rate (capacity) of the link connecting multiplexing node $i$ with previous node, and $\alpha_i^{TN}$ is the fraction of this link capacity that is allocated to a specific service. $\lambda_i^{TN}$ is the arrival rate of the service at this link of the transport network.

Downlink transport latency comprises both transit and propagation latencies, where the propagation latency ($l_{prop}^{TN}$) is symmetric for both uplink and downlink. Downlink traffic arriving at the transport network could originate from multiple UPFs. The aggregation of such heterogeneous flows results in a non-deterministic, non-Poisson input. In this case, the first link in the TN should be modeled as a GI/M/1 queue, where arrivals follow a general inter-arrival distribution with mean $1/\lambda_1^{TN}$ and variance $\sigma_{TN}^2$, while the service time ($T_{S_1}^{TN}$) remains exponential with rate $\mu_1^{TN}$. The utilization ratio is $\rho_1^{TN} = \lambda_1^{TN}/\mu_1^{TN}$, where $\rho_1^{TN} < 1$. In the GI/M/1 queue, the waiting time ($T_{W_1}^{TN}$) PDF is expressed as:

$$f_{T_{W_1}^{TN}}(t) = (1 - \beta_{TN})\delta(t) + \beta_{TN}\mu_1^{TN}(1 - \beta_{TN})e^{-\mu_1^{TN}(1-\beta_{TN})t}, t > 0, \tag{38}$$

where $\beta_{TN}$ is the unique positive root in the range (0,1) of the following equation:

$$\beta_{TN} = \phi_A(\mu_1^{TN} - \mu_1^{TN}\beta_{TN}), \tag{39}$$

with $\phi_A(s)$ being the LST of the inter-arrival PDF. The transit latency at the first link is $T_{TN_1} = T_{W_1}^{TN} + T_{S_1}^{TN}$, and its PDF is derived by the convolution $f_{T_{TN_1}}(t) = f_{T_{W_1}^{TN}}(t) * f_{T_{S_1}^{TN}}(t)$. Given the exponential service time $f_{T_{S_1}^{TN}}(t) = \mu_1^{TN}e^{-\mu_1^{TN}t}$, this convolution simplifies to a unified exponential PDF for transit latency:

$$f_{T_{TN_1}}(t) = (1 - \beta_{TN})\mu_1^{TN}e^{-(1-\beta_{TN})\mu_1^{TN}t}, t > 0. \tag{40}$$

Other links in the TN are modeled as M/M/1 queues since their inputs are coming from previous TN links (with M/M/1 or GI/M/1 models) and their service times are exponential. The transit latency at subsequent links $T_{TN_i}$ follows then an exponential distribution $f_{T_{TN_i}}(t) \sim Exp(\mu_i^{TN} - \lambda_i^{TN})$. At the output of the $n_{TN}$ multiplexing node, the transit latency is obtained as the following convolution:

$$f_{T_{TN}}(t) = f_{T_{TN_1}}(t) * f_{T_{TN_2}}(t) * \dots * f_{T_{TN_{n_{TN}}}}(t). \tag{41}$$

*C. Core Network*

The CN latency $l_{CN}$ is the sum of the propagation $l_{prop}^{CN}$ and transit latencies $T_{CN}$ as $l_{CN} = T_{CN} + l_{prop}^{CN}$, where the propagation latency is determined by the CN distance $d_{CN}$ and the propagation speed $v_{CN}$ for both uplink and downlink $l_{prop}^{CN} = \frac{d_{CN}}{v_{CN}}$.

To model the CN transit latency, we assume $n_{CN}$ serial nodes in the CN, with each node represented as a deterministic-service queue [21]. For the uplink, packets arrive from the TN, where arrivals follow a Poisson distribution following the M/M/1 queueing model adopted for the TN. Consequently, the queue at each CN node can be modeled as an M/D/1 queue [4] with arrivals following a Poisson process with rate $\lambda_i^{CN}$. The service time is deterministic with a fixed service rate $\mu_i^{CN}$. Based on the M/D/1 model, the expected transit latency including the service time (first term) and the waiting time (second term) for each CN node is:

$$\mathrm{E}\left[T_{CN_i}\right] = \frac{1}{\mu_i^{CN}} + \frac{\rho_i^{CN}}{2\mu_i^{CN}(1 - \rho_i^{CN})}, \tag{42}$$

where $\rho_i^{CN} = \lambda_i^{CN}/\mu_i^{CN}, \rho_i^{CN} < 1$. In addition, the transit latency PDF at each CN node can be approximated as

$$f_{T_{CN_i}}(t) = 2\lambda_i^{CN}(1 - \rho_i^{CN})e^{-2\lambda_i^{CN}\left(1-\rho_i^{CN}\right)(t-D)}, t \geq \mathrm{D}, \tag{43}$$

where D is the deterministic service time $D = \frac{1}{\mu_i^{CN}}$. Assuming the uplink traffic traverses $n_{CN}$ CN nodes, the CN transit latency results in a shifted Gamma (Erlang) distribution:

$$f_{T_{CN}}(t) = \frac{{\mu_i^{CN}}^{n_{CN}}(t-n_{CN}D)^{n_{CN}-1}}{(n_{CN}-1)!}e^{-\mu_i^{CN}(t-n_{CN}D)}, \tag{44}$$

where $t \geq n_{CN}D$.

In the downlink, the traffic arrives to the CN from the Internet, where inter-arrival times are highly variable and well described by a Weibull distribution [18]. In this case, the CN is modeled as a G/D/1 queue instead of M/D/1, still with deterministic service time but general arrival times. Let $A_{CN}$ denote the inter-arrival time. Its mean and variance are $\mathrm{E}[A_{CN}] = 1/\lambda_i^{CN}$ and $\mathrm{Var}[A_{CN}] = \sigma_{CN}^2$, respectively. The utilization is given by $\rho_i^{CN} = \lambda_i^{CN}/\mu_i^{CN}$, where $\rho_i^{CN} < 1$. The mean waiting time at each CN node is obtained from a Pollaczek–Khinchine–type expression:

$$E\left[T_{W_i}^{CN}\right] = \frac{\lambda_i^{CN}\sigma_{CN}^2}{2(1 - \rho_i^{CN})}. \tag{45}$$

The service time at each CN node is deterministic with its PDF equal to $f_{T_{S_i}^{CN}}(t) = \delta\left(t - \frac{1}{\mu_i^{CN}}\right)$. The PDF of the transit latency at each CN node is then:

$$f_{T_i^{CN}}(t) = f_{T_{W_i}^{CN}}\left(t - \frac{1}{\mu_i^{CN}}\right), \qquad t \geq \frac{1}{\mu_i^{CN}}. \tag{46}$$

For Internet-originated traffic, the waiting time PDF $f_{T_{W_i}^{CN}}(t)$ is approximated by an exponential distribution with mean $E\left[T_{W_i}^{CN}\right]$, leading to:

$$f_{T_i^{CN}}(t) \approx \frac{1}{E\left[T_{W_i}^{CN}\right]} exp\left(-\frac{t - \frac{1}{\mu_i^{CN}}}{E\left[T_{W_i}^{CN}\right]}\right), \quad t \geq \frac{1}{\mu_i^{CN}}. \tag{47}$$

At the output of the $n_{CN}$ core node, the transit latency is obtained as the following convolution:

$$f_{T_{CN}}(t) = f_{T_{CN_1}}(t) * f_{T_{CN_2}}(t) * \dots * f_{T_{CN_{n_{CN}}}}(t). \tag{48}$$

### *D. Internet*

In this study, the Internet latency in both the uplink and downlink directions is defined as the latency between the core network UPF and the cloud server. We model the Internet latency using the empirical study reported in [18], where the Internet latency is measured considering the transmission time, propagation time, and queuing times at intermediate routers/switches between a pair of nodes in the Internet path. The model characterizes the cumulative distribution function (CDF) of the round-trip time observed between pairs of Internet nodes as a Weibull distribution:

$$F(t) = 1 - exp\left(-\left(\frac{t}{0.0175}\right)^{1.87}\right), \tag{49}$$

where the scale parameter 0.0175 and the shape parameter 1.87 are obtained by fitting a Weibull distribution to the distribution curve presented in [18].

## VI. Model Validation

This section validates the proposed latency models. To this end, we benchmark the communication and computing latency models by comparing the estimated latencies of the various components with corresponding measurements reported in the literature, whenever available. It should be noted, however, that the scarcity of empirical measurements for an end-to-end continuum architecture of the type considered in this paper remains a recognized limitation.

### *A. Communication latency*

TABLE I reports the average, 90th, and 99th percentiles of the RAN latency estimates generated by our model. In our simulation setup, the RAN is configured with a 5 MHz channel bandwidth, a Signal-to-Noise Ratio (SNR) of 30 dB, and aperiodic traffic rates of 1 Mbits/s and 2 Mbits/s. We specifically compare our results with the aperiodic traffic data from [3] because the input traffic in our model is inherently aperiodic. Under aperiodic traffic conditions, the authors in [3] report average latencies in the range of 1.5 to 10 ms, 90th percentiles between 2 and 10 ms, and 99th percentiles ranging from 5 to 25 ms. As the table shows, our results fall perfectly within the same order of magnitude. It is important to note that the values in [3] were obtained using a larger 20 MHz bandwidth and a higher range of data rates. However, when considering the ratio of the data rate to the available bandwidth, both scenarios operate within a comparable range. Consequently, our model's estimates accurately capture the expected tail latencies under proportional network load conditions. The RAN latency estimates obtained with our model are also consistent with empirical 5G NR measurements reported in [22], which show average latencies ranging from 1 to 6 ms and 99th percentile latencies between 2 and 10 ms. These empirical results were obtained under industrial propagation conditions that create highly challenging Non-Line-of-Sight (NLOS) environments and were evaluated across different Time Division Duplexing (TDD) frame structures and link robustness (BLER target) configurations. Importantly, the measurements in [22] were conducted using a massive millimeter-wave (mmW) bandwidth of 800 MHz, accommodating data rates that are proportionally higher relative to this larger bandwidth. Considering that our model operates under a highly constrained 5 MHz channel bandwidth, our estimated average latencies and 99th percentiles are remarkably consistent and proportionally aligned with these empirical observations. Furthermore, it should be noted that while the packets in [22] were generated at equal intervals (periodic traffic), our model processes aperiodic traffic; this irregular packet generation naturally contributes to the latency variations observed in our percentiles.

TABLE I

RAN Latency (ms)

| **Traffic rates** | **Average** | **90th pctl** | **99th pctl** |
|---|---|---|---|
| 1 Mbits/s | 4.61 | 10.41 | 16.72 |
| 2 Mbits/s | 7.21 | 14.38 | 18.92 |

TABLE II reports the transport network (TN) latency values (average, 90th and 99th percentiles) obtained with the proposed model for different values of $\alpha_i^{TN}$. This is consistent with the TN latency values presented in [4] for various 5G deployment scenarios and capacity allocations, which reported values ranging from 0.402 ms to 10.279 ms. Furthermore, our results align with the empirical measurements reported in [23] from Vodafone's live 5G macro network deployed in London, where the transport network latency ranged from 1.68 ms to 5.6 ms depending on the routing configurations.

TABLE II

Transport Network Latency (ms)

| $\alpha_i^{TN}$ | **Average** | **90th pctl** | **99th pctl** |
|---|---|---|---|
| 0.01 | 0.77 | 0.80 | 0.85 |
| 0.001 | 1.75 | 3.05 | 5.35 |

The core network (CN) comprises both propagation latency and transit latencies. In our analysis, we consider an optical CN spanning 200 km, which results in a propagation latency of 2 ms (uplink and downlink). The average, 90th percentile and 99th percentile CN latencies obtained with the proposed model are 2, 2.001 and 2.001 ms, respectively. These results indicate that the CN latency is largely dominated by propagation latency, which is consistent with the experimental findings reported in [24] for a 5G deployment in Helsinki. Our CN latency estimates are also consistent with the values reported in [4] and [24] where core network latencies for MEC-supported deployments are similarly reported in the range of 2 to 3 ms.

Finally, it should be noted that Internet latency is modeled based on the empirical measurements reported in [18]. This latency component can be considered to have already been validated against real-world measurements.

*B. Computing latency*

TABLE III presents the average, 90th and 99th percentile computing latencies obtained with our local processor model for utilization levels ranging from 40% to 60%. These values closely align with the empirical measurements reported in [25], which evaluates the average service time for the Xapian, which is a well-known benchmark workload. [25] reports average values of approximately 1 ms, and 99th percentile latencies in the 5–8 ms range for different workload classes under a FCFS scheduling policy [25]. These values are consistent with the latency estimates produced by the proposed model and reported in TABLE III.

TABLE III
Local Processing Latency (ms)

| Usage Ratio | Average | 90th pctl | 99th pctl |
|---|---|---|---|
| 40% | 0.67 | 1.54 | 3.07 |
| 50% | 1 | 2.30 | 4.60 |
| 60% | 1.5 | 3.45 | 6.89 |

TABLE IV presents the average, 90th, and 99th percentile computing latencies obtained with the proposed edge and cloud computing latency models under different usage ratio. In our experimental setup, the edge node is configured with 5 processor cores at 4.3 GHz, and the results are evaluated across edge usage ratios ($\rho_e$) ranging from 7% to 21%. The latency estimates produced by the edge computing model closely align with the empirical measurements reported in [26]. For example, measurements obtained using the NVIDIA Jetson Orin Nano platform report average computing latencies ranging from 0.54 ms to a few milliseconds, depending on the middleware implementation. Similarly, in our experimental setup, the cloud node is configured with 25 processor cores at 4.3 GHz, and evaluated across cloud usage ratios ($\rho_c$) ranging from 1.4% to 4.2%. The cloud computing latencies obtained with the proposed model under these configurations typically range from 10 µs to 130 µs, which is consistent with the empirical measurements reported in [27] and [28]. For instance, [27] reports typical computing latencies in cloud datacenters to be between 60 µs and 100 µs. Furthermore, [28] shows that data access and processing latencies in lightweight cloud virtualization environments average around 150 µs.

TABLE IV
Edge and Cloud Computing Latencies

| Node | Usage ratio | Average | 90th pctl | 99th pctl |
|---|---|---|---|---|
| Edge | 7% | 0.44 ms | 1.01 ms | 2.01 ms |
| | 14% | 0.87 ms | 2.11 ms | 4.12 ms |
| | 21% | 1.31 ms | 3.15 ms | 6.31 ms |
| Cloud | 1.4% | 10 $\mu s$ | 20 $\mu s$ | 40 $\mu s$ |
| | 2.8% | 20 $\mu s$ | 40 $\mu s$ | 90 $\mu s$ |
| | 4.2% | 30 $\mu s$ | 60 $\mu s$ | 130 $\mu s$ |

*C. End-to-end latency*

TABLE V reports the end-to-end (E2E) latency values obtained with the proposed model for different traffic rates, defined by both the packet arrival rate and the individual packet size, when the computing is performed at the edge or the cloud. The RAN was configured with a 5 MHz bandwidth and a Signal-to-Noise Ratio (SNR) of 30 dB. For the computing environments, the edge node (5 cores at 4.3 GHz) was set to a usage ratio of 14%, and the cloud node (25 cores at 4.3 GHz) was set to a usage ratio of 2.8%.

In the edge computing case, the estimated E2E latencies are consistent with the empirical range of [7.8–20] ms reported in the literature [29]–[30]. Specifically, these empirical values were obtained in the 5G-MOBIX project, utilizing 5G NR NSA architectures where Multi-access Edge Computing servers were co-located with base stations (gNBs) and connected via dedicated high-capacity fiber links. In addition, the estimated E2E latencies, when computing is performed in the cloud, are consistent with the 53.8–150 ms range reported in the literature [29]–[30]. These empirical measurements, observed in the 5G-MOBIX project, reflect scenarios where user traffic was backhauled from 5G gNBs to centralized Evolved Packet Core sites and remote cloud data centers over multi-hop wide-area transport networks, accounting for standard Internet routing and core network processing overheads.

TABLE V
End-to-End Latencies (ms) for Edge and Cloud computing

| Node | Traffic rates | Average | 90th pctl | 99th pctl |
|---|---|---|---|---|
| Edge | 1 Mbits/s | 6.33 | 12.24 | 18.57 |
| | 2 Mbits/s | 9.32 | 16.56 | 21.55 |
| | 3 Mbits/s | 11.72 | 19.47 | 24.26 |
| Cloud | 1 Mbits/s | 40.44 | 58.31 | 75.91 |
| | 2 Mbits/s | 46.87 | 65.67 | 82.71 |
| | 3 Mbits/s | 52.24 | 71.72 | 88.69 |

## VII. Evaluation Scenario

We consider a vehicular scenario with different V2X service profiles. The computing demand of each service ranges from 1.5 to 6 MCycles per packet, while the latency deadlines vary between 10 ms and 100 ms. The packet rate is 1000 packets per second. A service can be executed locally using the vehicle's onboard computing resources. We assume that each service can use a local processing capacity of $\mu_l$= 12 GHz. Alternatively, the service can be offloaded to the edge or the cloud. In these cases, we assume that each service can use 5 processors ($m_e$) operating at $\mu_e = 4.3$ GHz at the edge, or 25 processors ($m_c$) operating at $\mu_c = 4.3$ GHz.

We consider a cellular connection with an average SINR ($\bar{\gamma}$) of 10 or 30 dB, and a transmission bandwidth BW of 5 MHz. Following a reference deployment proposed by the ITU-T Study Group in [19], the transport network consists of three levels ($n_{TN} = 3$) of multiplexing nodes (M1, M2 and M3) connecting a number of gNBs to the CN. The distance between the gNB and the M1 nodes is $d_{TN}(M_1)$=3 km, the distance between the M1 and M2 nodes is $d_{TN}(M_2)$ =12 km, and the distance between M2 and M3 nodes is $d_{TN}(M_3) = 60$ km. We consider link capacities of $C_1^{TN}$ =10 Gb/s for the gNB-M1 links, $C_2^{TN}$ =300 Gb/s for the M1-M2 links, and $C_3^{TN}$ =6 Tb/s for the M2-M3 links [4]. We assume $\alpha_1^{TN} = \alpha_2^{TN} = \alpha_3^{TN} = 0.01$ for different levels of transport network. Following ITU-T Study Group reports, we consider a CN distance of $d_{CN}$=200 km with a link capacity between the CN nodes of $\mu^{CN} = 6$ Tb/s [4].

Without loss of generality, we consider in-vehicle network traffic that can be executed locally or uploaded to the edge or cloud. This traffic may experience jitter due to synchronization errors, contention on shared resources, or data aggregation from multiple sources, among others. To characterize this jitter, we utilize the in-vehicle network dataset available in [31], which captures traffic generated by different embedded sensors, electronic control units (ECUs) and CPUs.

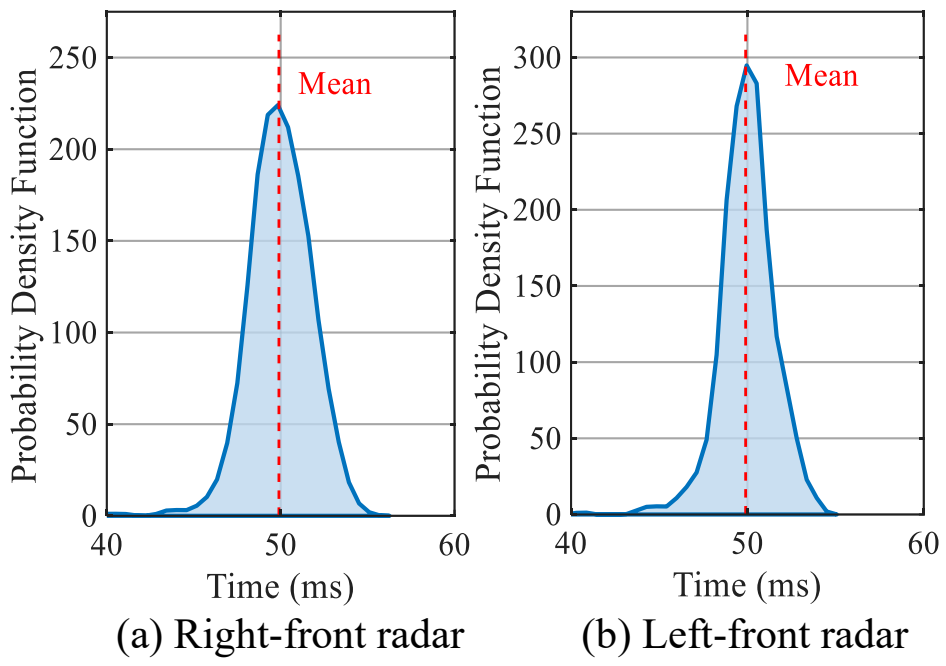

**Fig. 2.** Distribution of packet arrival-times.

Fig. 2 depicts the distribution of packet arrival times of traffics form right-front radar (Fig. 2(a)) and left-front radar (Fig. 2(b)) [31] which is measured in milliseconds. The distribution follows a Gaussian distribution, which we use to model the jitter $J_k$. Accordingly, the arrival time $A_k$ of the $k^{th}$ packet is:

$$A_k = T_{ideal,k} + J_k, \quad (50)$$

where $T_{ideal,k}$ denotes the ideal packet arrival time and $J_k$ follows a Gaussian distribution $J_k \sim \mathbb{N}(\mu_J, \sigma_J^2)$, with $\mu_J$ and $\sigma_J^2$ representing the mean and variance of the jitter, respectively. For a sequence of packets with a constant inter-arrival time $T$, the inter-arrival time $\tau_k$ is given by:

$$\tau_k = A_{k+1} - A_k = T + (J_{k+1} - J_k). \quad (51)$$

Since $J_{k+1}$ and $J_k$ are independent Gaussian variables, their difference is also Gaussian with variance $2\sigma_J^2$. Consequently, the inter-arrival time $\tau$ follows a Gaussian distribution as $\tau \sim \mathbb{N}(T, 2\sigma_J^2)$. The corresponding Laplace-Stieltjes Transform (LST) of the inter-arrival time is:

$$\mathcal{L}_\tau(s) = e^{-sT + s^2\sigma_J^2}. \quad (52)$$

Equation in (52) shows that the variance of the inter-arrival times is directly determined by the jitter variance $2\sigma_J^2$, providing a closed-form analytical solution for the inter-arrival distribution. To facilitate a better comparison across different scenarios, a normalized jitter is employed in our evaluations, which is defined as $C_J = \frac{\sigma_J}{\mu_J}$.

To model traffic variability in the in-vehicle network, we assume a bounded worst-case scenario in which the packet arrival rate follows a uniform distribution. Unlike unbounded distributions, which can theoretically produce arbitrarily large bursts, a uniform distribution ensures that the instantaneous traffic load remains within defined physical limits. The mean arrival rate is denoted by $\lambda_V$, and the variability is controlled through the standard deviation $\sigma_V$. We map the $C_V = \frac{\sigma_V}{\mu_V}$ ratio to the traffic variability. This approach allows us to model bursty traffic while respecting upper-bound limits. By sweeping $C_V$, we evaluate the impact of traffic variability on the end-to-end latency.

## VIII. Impact of Arrival-Time Jitter

This section evaluates the impact of arrival-times jitter on the end-to-end latency and the probability of meeting latency deadlines across the continuum. Fig. 3 shows the probability of meeting the latency deadlines when services are executed locally on the vehicle, at the edge, or in the cloud. The results are presented as a function of the arrival-time jitter $C_J$ for services with different computing demands and latency deadlines. The results in Fig. 3 are obtained assuming an average cellular SINR $\bar{\gamma}$ of 30 dB. Fig. 3(a) shows that, when the jitter is low ($C_J \approx 0$), the latency deadlines are met for most services when they are executed locally. As the jitter increases, its impact depends on the service latency requirement, with services having the shortest deadlines being the most sensitive to the jitter. This sensitivity becomes more evident when services are offloaded to the edge (Fig. 3(b)) or the cloud (Fig. 3(c)), with the highest sensitivity observed for cloud execution due to the additional communication latency. For example, the service with a 30 ms latency deadline and the highest computing demand cannot be satisfactorily executed at the edge, whereas it can when processed locally. Similarly, all services with a 30 ms deadline, regardless of their computing demand, exhibit a lower probability of meeting their deadline in the presence of jitter when executed in the cloud that when processed locally or at the edge. Fig. 3 also shows that services with the longest latency deadlines remain resilient to high levels of jitter when executed locally. However, their resilience decreases when the services are offloaded to the cloud, where the additional communication latency significantly reduces the available latency budget.

Fig. 4 plots the 99th percentile end-to-end latency as a function of the jitter for the different services, thereby characterizing the system's tail-latency behavior. The figure shows that the 99th percentile latency increases with the jitter and reaches its highest values when services are offloaded to the cloud. As the jitter increases, the 99th percentile latency augments for all services. However, the increase is more pronounced for services with higher computing demands, demonstrating that jitter amplifies tail latency more significantly for compute-intensive workloads. This effect is consistent regardless of whether services are executed locally, at the edge, or in the cloud. It occurs because jitter causes packets to arrive at the computing queue at irregular times, increasing queues length and waiting times for the services with the highest computing demand. Consequently, the tasks for these services remain in the system for longer periods, leading to a larger increase in tail latency.

Fig. 3 and Fig. 4 show that the impact of the tail-latency growth on the probability of completing a task before its latency deadline depends on both the service latency deadline and where tasks are executed. For example, Fig. 4(a) shows that the jitter significantly augments the 99th percentile latency for the service with a 100 ms deadline and for the service with a 30 ms deadline and a 6 Mcycles computing demand when they are executed locally. Despite this increase, the probability of meeting their latency deadlines remains essentially unchanged (Fig. 3(a)). In contrast, the service with a 10 ms deadline experiences a smaller increase in tail latency, yet this increase is sufficient to noticeably reduce its probability of meeting the deadline because of its more stringent latency requirement. A similar trend is observed when services are offloaded to the edge or cloud. However, the impact of tail-latency growth on the probability of meeting the latency deadline is higher in these cases because the communication latency consumes part of the available latency budget. This effect is illustrated by the services with a 30 ms deadline and computing demands of 1.5 MCycles and 6 MCycles. Although both services experience an increase in tail latency with the jitter (Fig. 4), this increase does not impact deadline satisfaction when they are executed locally (Fig. 3(a)), whereas it reduces the probability of meeting the deadline when they are offloaded to the edge (Fig. 3(b)) or to the cloud (Fig. 3(c)).

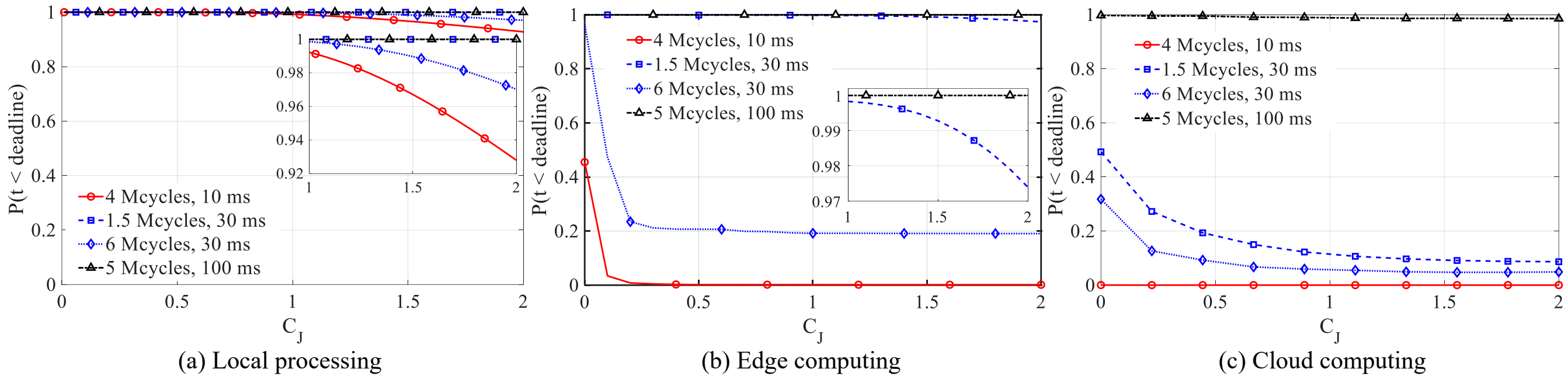

**Fig. 3.** Probability of meeting the latency deadline as a function of the jitter.

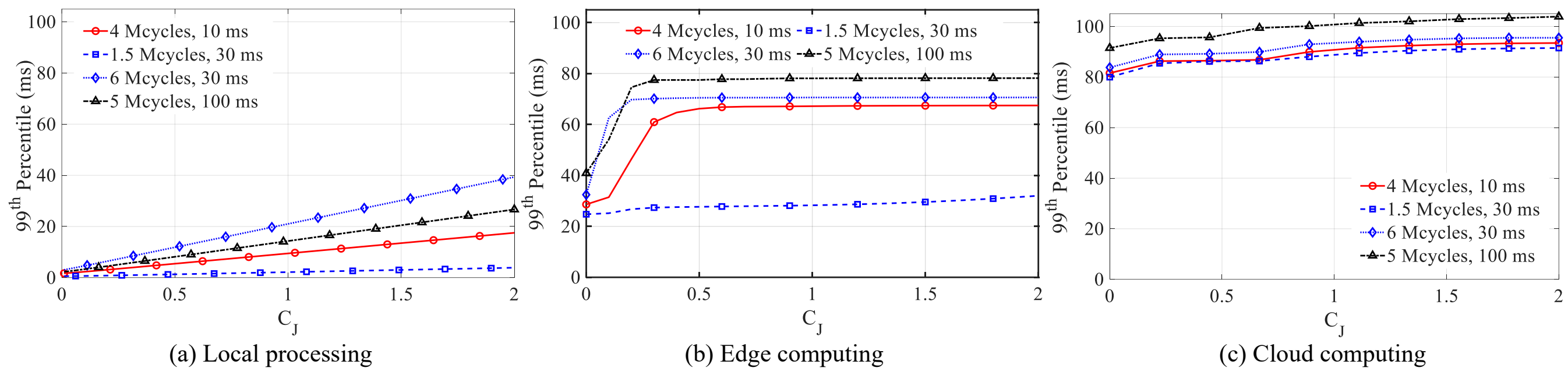

**Fig. 4.** $99^{th}$ percentile latency as a function of the jitter.

Results in Fig. 3 suggest that services with stringent latency deadlines should be executed locally to avoid the additional communication latency incurred when offloaded to the edge or the cloud. This is the case, for example, for services with latency deadlines of 10 ms or 30 ms. However, jitter can make it challenging to satisfy these deadlines even when services are executed locally, particularly under high processor workloads. This effect is illustrated in Fig. 5, which shows the impact of jitter on the probability of meeting the latency deadline under different local processing workloads. The figure compares the performance when the available local processing capacity ($\mu_l$= 12 GHz) is completely free, and when 25% and 45% of the local processing capacity is occupied. The results show that the impact of jitter becomes increasingly significant for services with stringent latency requirements as the processor workload grows. Under limited available processing capacity, jitter increases queueing delays, causing tasks to remain longer in the processor queue and reducing the probability of meeting the latency deadline. In contrast, services with more relaxed deadlines are only marginally affected by jitter, even under higher processor workloads. Although Fig. 5 presents results for locally processed services, the same interaction between jitter and processing workload is also observed for services executed at the edge or in the cloud. Similar trends are also observed when different service types are executed with different processor occupancy levels.

Fig. 5 shows that services with more relaxed deadlines (100 ms in our analysis) can be satisfactorily executed locally despite the presence of jitter when the processor workload is low. However, as both the processor workload and the jitter increase, the probability of meeting the latency deadline drops because the larger queueing delays consume an increasing fraction of the available latency budget. An alternative is to offload these services to the edge or the cloud since their relaxed latency deadline can potentially accommodate the additional communication latency. However, the communication latency depends on the quality of the cellular connection. Fig. 6 compares the probability of meeting the latency deadline for a service with a 100 ms deadline when offloaded to the edge or the cloud under good (30 dB) and poor (10 dB) cellular link quality conditions. The results show that offloading the service to the edge achieves a considerably higher probability of meeting the latency deadline than local execution under high processor workload (Fig. 5 with 45% occupancy), regardless of the jitter level. This improvement is observed under both good and poor cellular link quality conditions because the communication latency only includes the RAN latency when the service is offloaded to the edge. Although the RAN latency increases as the cellular link quality deteriorates, the latency budget available for services with relaxed deadlines remains sufficient to absorb this increase. This is not the case when the service is offloaded to the cloud. In this case, the communication latency also includes the transport, core, and Internet latencies, which further reduce the available latency budget. As a result, cloud execution becomes significantly more sensitive to both cellular link quality and jitter. Fig. 6 shows that, under low jitter and good cellular conditions, cloud execution can still achieve a high probability of meeting the latency deadline. However, as the cellular link quality deteriorates and/or the jitter increases, this probability decreases substantially. Under these conditions, executing the service locally may outperform cloud execution, even when the local processor operates under high workload (Fig. 5). These results highlight the importance of controlling jitter to prevent performance degradation when offloading services to the cloud.

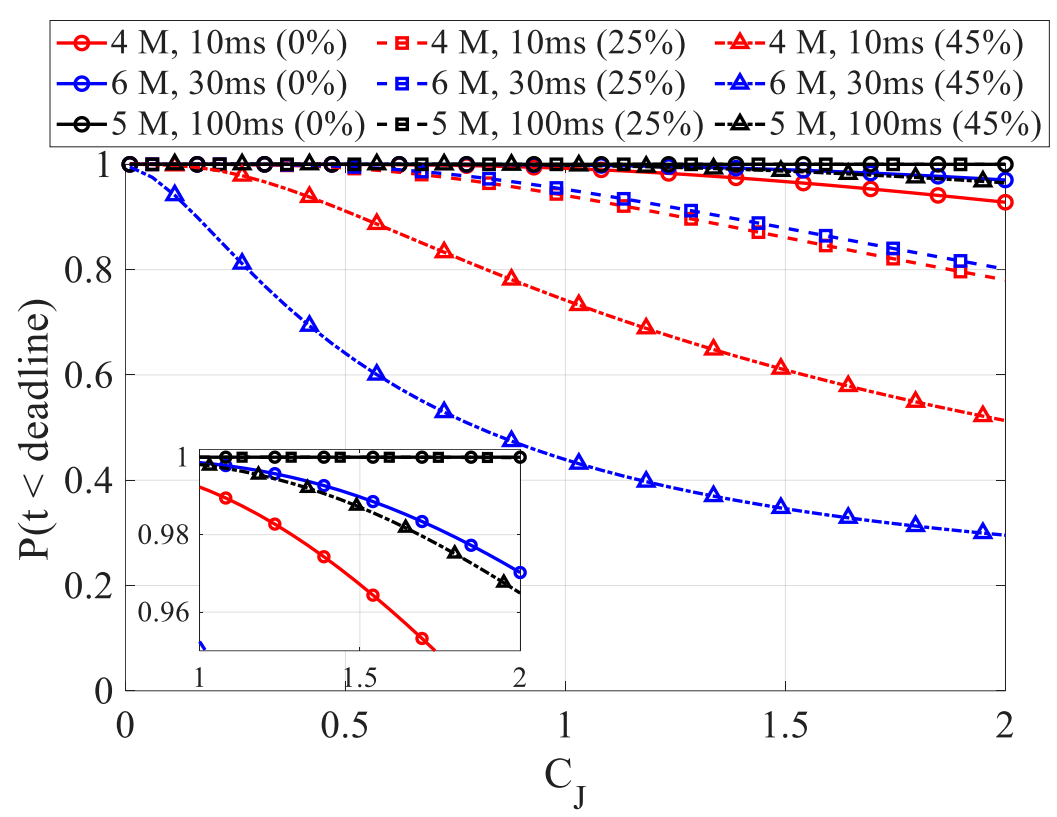

**Fig. 5.** Probability of meeting the latency deadline as a function of the jitter and the local processing workload.

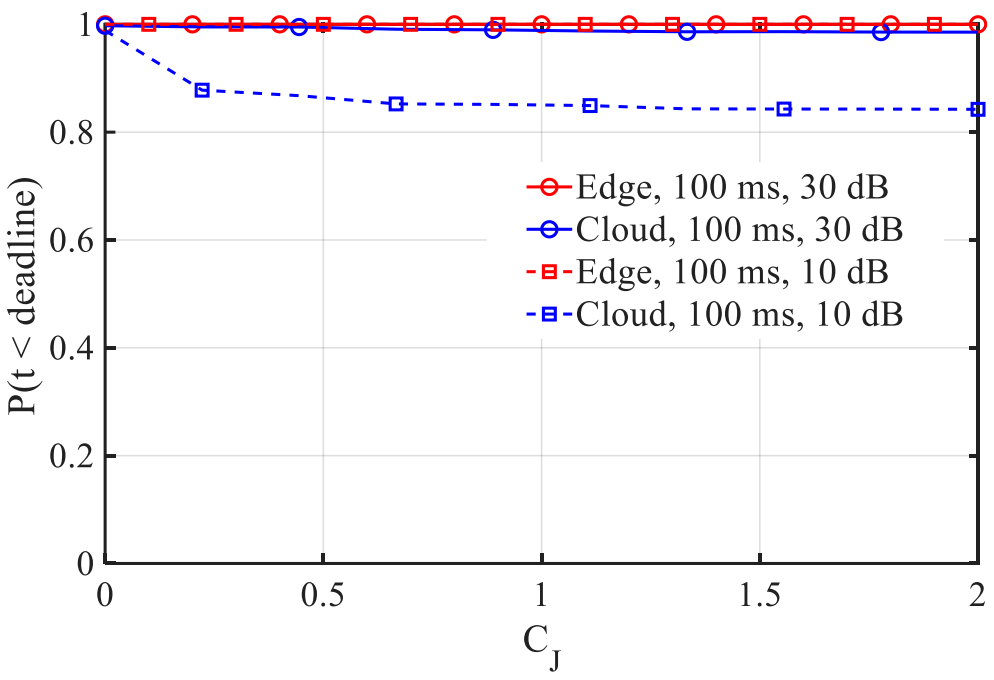

**Fig. 6.** Probability of meeting the latency deadline as a function of the jitter and the link quality conditions.

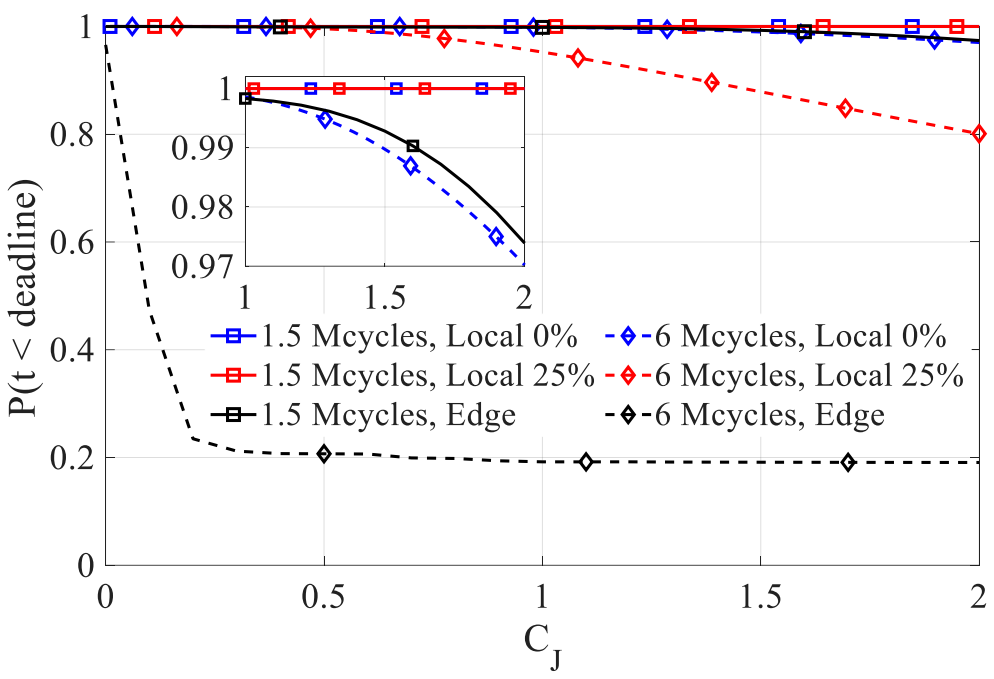

**Fig. 7.** Probability of meeting the latency deadline as a function of the jitter and the local processor occupancy.

The combined effects of jitter and cellular link quality become even more pronounced for services with shorter latency deadlines, while their impact also depends on the service computing demand. This behavior is illustrated in Fig. 7, which compares the effect of jitter for a service with a 30 ms deadline and computing demands of 1.5 Mcycles and 6 Mcycles. The results correspond to an average cellular link quality of 30 dB and consider local execution (with zero and 25% occupancy) as well as edge execution. The figure shows that, regardless of the computing demand, the probability of meeting the latency deadline decreases as the jitter increases when the local processor workload grows. For services with a low computing demand (1.5 Mcycles), edge execution is less affected by jitter. However, this advantage disappears when the computing demand increases to 6 Mcycles. In this case, Fig. 7 shows that offloading the service to the edge provides no performance benefit (and cloud execution performs even worse), since the probability of meeting the latency deadline deteriorates more rapidly with increasing jitter than for local execution. This behavior results from the reduced latency budget available after accounting for the communication latency and the longer computing latency associated with the higher computing demand.

## IX. Impact of Traffic Variability

This section evaluates the impact of traffic variability. Fig. 8 shows the probability of meeting the latency deadline when services are executed locally, at the edge, or in the cloud. The results are presented as a function of the normalized rate variation $C_V = \frac{\sigma_V}{\lambda_V}$, where $\lambda_V$ represents the average traffic rate and $\sigma_V$ indicates its variation. The results are reported for services with different latency deadlines and computing demands, assuming an average cellular SINR $\bar{\gamma}$ of 30 dB. A comparison of Fig. 3 and Fig. 8 shows that traffic variability impacts the probability of meeting the latency deadline in a manner similar to jitter, although its impact is generally more pronounced. As the traffic variability ($C_V$) increases, the probability of meeting the latency deadline decreases, with the degradation being more severe for services with more stringent latency deadlines and higher computing demands. The sensitivity to traffic variability also increases when services are offloaded to the edge or the cloud, even for the services with the more relaxed latency deadlines, due to the additional communication latency.

We also analyzed the impact of traffic variability on the 99th-percentile end-to-end latency to characterize the system's tail-latency behavior. The corresponding results are omitted for brevity, as they exhibit trends similar to those observed for jitter. In particular, the 99th-percentile latency increases with traffic variability, with the largest increases observed for services offloaded to the cloud and for services with higher computing demands. Consistent with the jitter analysis, the increase in tail latency does not necessarily translate into a lower probability of meeting the latency deadline. The degradation is primarily observed for services with stringent latency requirements even when the increase in tail latency is modest. It is also observed, to a lesser extent, for services with intermediate latency deadlines when offloaded to the edge or the cloud since the communication latency reduces the available latency budget. Services with more relaxed latency deadlines remain the least affected despite exhibiting the highest tail-latency values.

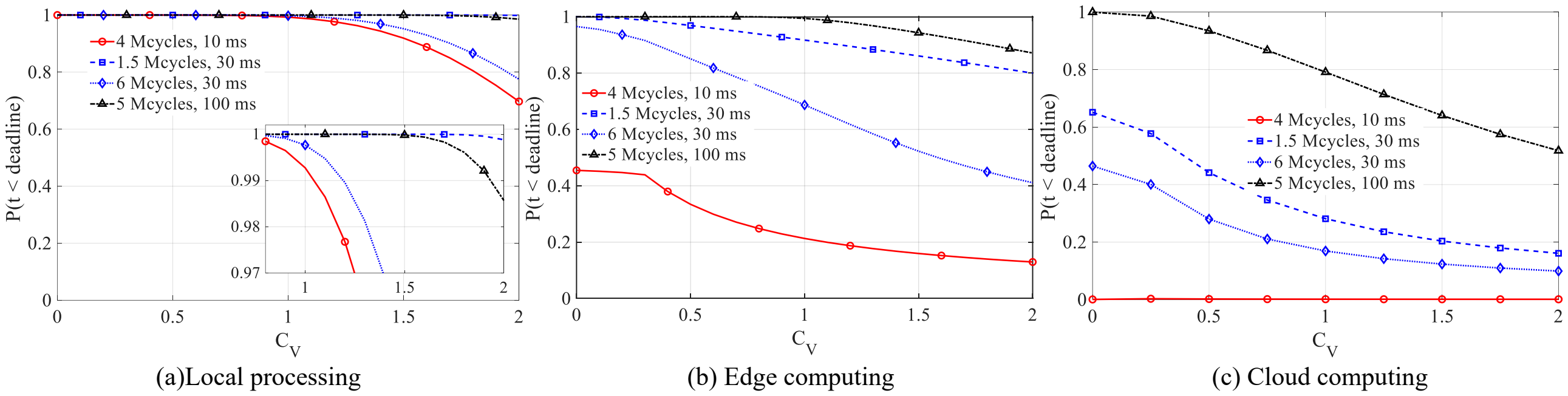

**Fig. 8.** Probability of meeting deadlines with respect to traffic variability in different computing nodes.

We have also analyzed the impact of traffic variability as a function of the processor workload, cellular quality conditions, and service computing demand. The corresponding results are omitted for brevity, as they lead to the same qualitative conclusions as those obtained for jitter. In particular, increasing traffic variability reduces the probability of meeting the latency deadline, with the degradation becoming more pronounced for services with stringent latency requirements and higher computing demands. Services offloaded to the edge or the cloud exhibit greater sensitivity to traffic variability due to the additional communication latency. This impact is more severe than that of jitter because traffic variability increases the amount of data transmitted over the communication network, thereby increasing the communication latency, reducing the available latency budget, and consequently lowering the probability of meeting the latency deadline. Similarly, the impact of traffic variability becomes more pronounced as the processor workload increases, since traffic variability also leads to longer processing queues under limited computing capacity.

## X. Conclusion

This paper presents an end-to-end latency modeling framework for the IoT–edge–cloud continuum that jointly captures computing and communication latency across the radio access network, transport network, core network, Internet, and computing infrastructure. By incorporating queueing-based stochastic analysis, the proposed framework goes beyond average latency metrics and is capable of characterizing the complete latency distribution, including tail latency and the probability of meeting service deadlines. The model is released as open source for the research community, and is used in this work to analyze the impact of jitter and traffic variability on the ability to support deterministic service profiles across the continuum. The results demonstrate that meeting service latency deadlines and supporting deterministic service levels are fundamentally determined by the interplay between each service's latency deadline and the stochastic traffic variability experienced along the communication–computing chain. Services with stringent latency deadlines are highly sensitive to both jitter and traffic variability, making local processing the preferred option, as the additional communication latency introduced by edge and cloud computing leaves insufficient latency budget to tolerate latency variability. In contrast, services with more relaxed latency deadlines are considerably more resilient to jitter and traffic variability when executed locally or at the edge, despite increases in average and tail latency, because their larger latency budgets can accommodate the additional uncertainty. The gains achieved through edge offloading are particularly significant when the local processor operates under high workload and the cellular link quality is good, thereby limiting the impact of communication latency on the available latency budget. Cloud computing, however, is considerably more sensitive to traffic variability because the additional communication latency further reduces the available latency budget. These findings highlight that effective service offloading decisions must jointly consider service latency requirements, computing and communication conditions, and the different sources of latency variability in order to reliably support deterministic service levels across the IoT–edge–cloud continuum.